\documentclass[prl,aps,groupedaddress,amsmath,amssymb,twocolumn]{revtex4}
\usepackage{graphicx}

\newcommand{\torate}[1]{\overset{#1}{\to}}
\newcommand{\tofromrate}[2]{\underset{#2}{\overset{#1}{\rightleftharpoons}}}
\newcommand{\molecule}[1]{{\mathrm{#1}}}
\newcommand{\MS}{\molecule{S}}
\newcommand{\W} {\molecule{W}}
\newcommand{\X}{\molecule{X}}

\begin{document}
\title{Mutual information between in- and output trajectories of biochemical
networks}
\author{Filipe Tostevin}
\affiliation{FOM Institute for Atomic and Molecular Physics, Kruislaan 407, 
1098 SJ Amsterdam, The Netherlands}
\author{Pieter Rein ten Wolde}
\affiliation{FOM Institute for Atomic and Molecular Physics, Kruislaan 407, 
1098 SJ Amsterdam, The Netherlands}
\date{April 9th, 2009}

\begin{abstract}
Biochemical networks can respond to temporal characteristics of time-varying 
signals. To understand how reliably biochemical networks can transmit
information we must consider how an input signal as a function of time---the
input trajectory---can be mapped onto an output trajectory. Here we estimate the
mutual information between in- and output trajectories using a Gaussian model.
We study how reliably the chemotaxis network of {\em E. coli} can transmit
information on the ligand concentration to the flagellar motor, and find the
input power spectrum that maximizes the information transmission rate.
\end{abstract}


\maketitle

Cells continually have to respond to a wide range of intra- and extracellular
signals. These signals have to be detected, encoded, transmitted and decoded by
biochemical networks. In the absence of biochemical noise, a particular input
signal will lead to a unique output signal, allowing the cell to respond
appropriately. Recent experiments, however, have vividly demonstrated that
biochemical networks can be highly stochastic \cite{Elowitz02}, and a key
question is therefore how reliably biochemical networks can transmit information
in the presence of noise.

To address this question, we must recognize that the message may be contained in
the {\em temporal dynamics} of the input signal. A well-known example is
bacterial chemotaxis, where the concentration of the intracellular messenger
protein depends not on the current ligand concentration, but rather on whether
this concentration has changed in the recent past \cite{Block83}---the response
of the network thus depends on the {\em history} of the input signal. Moreover,
the input signal may be encoded into the temporal dynamics of the signal
transduction pathway. For example, stimulation of the rat PC-12 system with a
neuronal growth factor gives rise to a sustained response of the Raf-Mek-Erk
pathway, while stimulation with an epidermal growth factor leads to a transient
response \cite{Marshall95}. In all these cases, the message is encoded not in
the concentration of some chemical species at a specific moment in time, but
rather in its concentration as a function of time. Importantly, whether the
processing network can reliably respond to a signal depends not only on the
instantaneous value of the signal, but also on the time scale over which it
changes. In general, the in- and output signals of biochemical networks are
time-continuous signals with non-zero correlation times. To understand how
reliably biochemical networks can transmit information, we need to know how
accurately an input signal as a function of time---the input {\em
trajectory}---can be mapped onto an output trajectory. In this article, we take
an information theoretic approach to this question.

A natural measure for the quality of information transmission is the mutual
information between the input signal $I$ and the network response $O$, given by 
$M(I,O)=H(O)-H(O|I)$ \cite{Shannon1948}. Here, $H(O)\equiv-\int dOp(O)\log
p(O)$, with $p(O)$ the probability distribution of $O$, is the information
entropy of the output $O$; $H(O|I)\equiv-\int dIp(I)\int dOp(O|I)\log p(O|I)$
is the average (over inputs $I$) information entropy of $O$ given $I$, with
$p(O|I)$ the conditional probability distribution of $O$ given $I$. Recently,
the mutual information between the {\em instantaneous} values of the in- and
output signals of biochemical networks has been investigated \cite{Detwiler00,
Ziv07}, although in these studies the temporal correlations in the input signals
were ignored. Here we investigate the mutual information between in- and output
trajectories.

{\em Mutual information between trajectories}---We consider a biochemical
network in steady state which has one input species S with copy number $S$ and
one output species X with copy number $X$. The mutual information between in-
and output trajectories is found by taking the possible input and output signals
$I$ and $O$ to be the possible trajectories $S(t)$ and $X(t)$:
\begin{equation}
M(S,X)=\int{\cal D} S(t)\int{\cal D}X(t)p(S(t),X(t))
	\log\frac{p(S(t),X(t))}{p(S(t))p(X(t))}. \label{eq:I}
\end{equation}
Calculating the mutual information between trajectories is in general a
formidable task, given the high-dimensionality of the trajectory space. However,
for a Gaussian model, which we will employ here, the mutual information can be
obtained analytically.

In this Gaussian model, it is assumed that the input signal consists of small
temporal variations around some steady-state value, obeying Gaussian statistics.
This limits our approach, but seems a reasonable simplification given that the
input statistics have not been measured for most, if not all, biological
systems. Moreover, we assume that the coupling between the components can be
linearized and that the intrinsic noise is small and Gaussian, according to the
linear-noise approximation \cite{VanKampen}; recent modeling studies have shown
this gives a good description of the noise properties of a large class of
biochemical networks, even when the copy numbers are as low as ten
\cite{TanaseNicola06,Ziv07}.  Under these assumptions the joint probability
distribution of the in- and output signals is described by a multivariate
Gaussian,
\begin{equation}
p({\bf v})=\frac{1}{(2\pi)^N|{\bf Z}|^{1/2}} 
	\exp\left(-\frac{1}{2}{\bf v}^{\rm T}{\bf Z}^{-1}{\bf v}\right).
	\label{eq:p_gaussian}
\end{equation}
The vector ${\bf v}\equiv({\bf s},{\bf x})$, with ${\bf s}=(s(t_1),
s(t_2),\dots,s(t_N))$ constructed from the input signal sampled at times $t=t_1,
\dots,t_N$, and ${\bf x}=(x(t_1),x(t_2),\dots,x(t_N))$; $s(t)$ and $x(t)$ are
the deviations of $S$ and $X$ away from their steady-state values, $\langle
S\rangle$ and $\langle X\rangle$, respectively. The $2N \times 2N$ covariance
matrix ${\bf Z}$ has the form
\begin{equation}
{\bf Z}=\begin{pmatrix} {\bf C}^{ss} & {\bf C}^{xs}\\
			{\bf C}^{sx} & {\bf C}^{xx}\end{pmatrix}, \label{eq:Z}
\end{equation}
where ${\bf C}^{\alpha \beta}$ is an $N\times N$ matrix with elements
${\bf C}^{\alpha\beta}_{ij}=C_{\alpha\beta}(t_i-t_j)=\langle\alpha(t_i)
\beta(t_j)\rangle$.
In the limit that the in- and output signals are time-continuous, the mutual
information rate between the in- and output trajectories $R({\bf s},{\bf x})=
\lim_{T\rightarrow\infty}M({\bf s},{\bf x})/T$ is given by \cite{Munakata}
\begin{eqnarray}
R({\bf s},{\bf x})=-\frac{1}{4\pi}\int_{-\infty}^{\infty}d\omega\ln
	\left[1-\frac{|S_{\rm sx}(\omega)|^2}{S_{\rm ss}(\omega)S_{\rm xx}
	(\omega)}\right], \label{eq:I_S_omega}
\end{eqnarray}
where the power spectrum $S_{\alpha \beta}(\omega)$ is the Fourier transform of
$C_{\alpha\beta}(t)$. Measuring the output signal as a function of time, $R({\bf
s},{\bf x})$ is the rate at which the information on the input trajectory
increases with time; importantly, $R({\bf s},{\bf x})$ takes into account
temporal correlations in the in- and output signal. We emphasize that Eq.
\ref{eq:I_S_omega} is exact only for linear systems with Gaussian statistics.
Importantly, however, Eq. \ref{eq:I_S_omega} can also be applied to systems
which do not obey Gaussian statistics and to non-linear systems; in these cases
it provides a lower bound on the channel capacity of the network \cite{Mitra01}.

A biochemical network differs from a channel in telecommunication or
electronics, in that the reaction that detects the input signal may introduce
correlations between the signal and the intrinsic noise of the reactions that
constitute the processing network \cite{TanaseNicola06}; these correlations are
a consequence of the molecular character of the components and thus unique to
(bio)chemical systems. If the detection reaction does not introduce
correlations, then the power spectrum of the output signal, $S_{\rm xx}
(\omega)$, is given by the spectral addition rule \cite{TanaseNicola06}:
\begin{equation}
S_{\rm xx}(\omega)=N(\omega)+g^2(\omega)S_{\rm ss}(\omega). \label{eq:S_xx}
\end{equation}
Here, $N(\omega)$ is the intrinsic noise of the processing network, $S_{\rm
ss}(\omega)$ is the power spectrum of the input signal, and $g^2(\omega)=
|S_{\rm sx}(\omega)|^2/S_{\rm ss}(\omega)^2$ is the frequency-dependent gain.
Identifying the spectrum of the transmitted signal as $P(\omega) = g^2(\omega)
S_{\rm ss}(\omega)$, Eq. \ref{eq:I_S_omega} can be rewritten as 
\begin{equation}
R({\bf s},{\bf x})=\frac{1}{4\pi}\int_{-\infty}^{\infty} d\omega
	\ln\left[1+\frac{P(\omega)}{N(\omega)}\right], \label{eq:I_omega}
\end{equation}
a well-known result for a time-continuous Gaussian channel \cite{Fano}. When the
detection reaction does introduce correlations between the input signal and the
noise of the processing network, one can still define $g^2(\omega)$ and 
$P(\omega)$ as above and apply Eq. \ref{eq:I_omega}. However, in this case
$N(\omega)$ and $g^2(\omega)$ are not intrinsic properties of the network, but
also depend on the statistics of the input signal.

{\em Network motifs}---The three elementary detection motifs shown in Table
\ref{tab:1} \cite{TanaseNicola06} illustrate a number of characteristics of the
transmission of trajectories. As a simple example of a time-continuous input
signal with a non-zero correlation time, we take the dynamics of $S$ to be a
Poissonian birth-and-death process; for large copy numbers, this gives
distributions that are approximately Gaussian.

\begin{table*}[tb]
\begin{tabular*}{0.9\textwidth}{@{\extracolsep{\fill}}ccccc}
Scheme & Reaction & $g^2(\omega)$ & $N(\omega)$&
$|S_{\rm sx}(\omega)|^2/S_{\rm ss}(\omega)S_{\rm xx}(\omega)$
\\[3pt]
\hline\\[-10pt]
(I) & $\MS + \W \tofromrate{\nu=k_f\W}{\mu} \X$ &
$\left[\frac{\nu(\mu+\nu+\lambda)}{\omega^2+(\mu+\nu)^2+\nu\lambda}\right]^2$ &
$\frac{2\nu\langle{S}\rangle}{\omega^2+(\mu+\nu)^2+\nu\lambda}$&
$\frac{\nu\lambda(\mu+\nu+\lambda)^2}{
	\left[\omega^2+(\mu+\nu)^2+\nu\lambda\right]
	\left[\omega^2+\lambda(\lambda+\nu)\right]}$
\\[5pt]
\hline\\[-10pt]
(II) & $\MS\torate{\nu}\X\torate{\mu}\emptyset$ &
$\frac{\nu^2\left[\omega^2+(\lambda+\nu)^2\right]}
{4(\lambda+\nu)^2(\omega^2+\mu^2)}$ &
$\frac{\nu\langle{S}\rangle(4\lambda+3\nu)}{2(\lambda+\nu)(\omega^2+\mu^2)}$ &
$\frac{\nu}{4(\lambda+\nu)}$ 
\\[5pt]
\hline\\[-10pt]
(III) &$\MS\torate{\nu}\MS+\X$, $\X\torate{\mu}\emptyset$ &
$\frac{\nu^2}{\omega^2+\mu^2}$ & 
$\frac{2\nu\langle{S}\rangle}{\omega^2+\mu^2}$ & 
$\frac{\nu\lambda}{\omega^2+\lambda(\lambda+\nu)}$
\\[5pt]
\hline
\end{tabular*}
\caption{Three elementary detection motifs. The input signal is modeled via
	 $\emptyset\overset{k}{\longrightarrow}{\mathrm{S}}$ and 
	 ${\mathrm{S}}\overset{\lambda}{\longrightarrow}\emptyset$. }
\label{tab:1}
\end{table*}

Motif I describes the reversible binding between, for example, a ligand and a
receptor, or an enzyme and its substrate. For this motif only we take the input
signal to be the total number of both bound and unbound molecules $S_T(t)
=S(t)+X(t)$. We find that this motif acts as a low-pass filter for information.
Specifically, the gain-to-noise ratio $g^2(\omega) /N(\omega)$, which determines
how accurately an input signal at frequency $\omega$ can be transmitted, is
approximately constant at low frequencies but decays as $\omega^{-2}$ for high
frequencies. Since input signals of biochemical networks are commonly detected
via this motif, this result suggests that high-frequency input signals are
typically not propagated reliably.

Motif II describes the scenario in which the signaling molecule is deactivated
upon detection. An important example is activation of membrane receptors by
ligand binding followed by endocytosis. If the input signal is a Poissonian
birth-and-death process, the mutual information between instantaneous values of
$S$ and $X$ is zero
\footnote{In the Gaussian model of Eq. \ref{eq:p_gaussian}, the mutual
          information between instantaneous values of $s$ and $x$ is $M^{\rm
          ins}(s,x)=-\frac{1}{2}\ln\left[1-\sigma_{\rm sx}^4/(\sigma_{\rm ss}^2
          \sigma_{\rm xx}^2)\right]$, where $\sigma^2_{\rm ss}=\langle
          s^2\rangle$, $\sigma^2_{\rm xx}=\langle x^2\rangle$, and
          $\sigma^2_{\rm sx} =\langle sx\rangle$.}---$X$ gives no
information about the current value of $S$. Indeed, to understand how cells can
use this motif to transmit information, we must consider the mutual information
between in- and output trajectories. Interestingly, for this motif $N(\omega)$
vanishes at high frequencies, while $g^2(\omega)$ approaches a constant value;
the gain-to-noise ratio thus diverges at high frequencies, meaning that this
motif can reliably transmit rapidly varying input signals 
\footnote{$R(S,X)$ for motif II diverges if the integral in
          Eq. \ref{eq:I_S_omega} is taken to infinity. In reality the integral
          is bounded because a reaction cannot happen infinitely fast.}.

Motif III is a coarse-grained model for enzymatic reactions or gene activation;
the enzyme-substrate or transcription-factor-DNA binding reaction, respectively,
has been integrated out. For this motif, which in contrast to the other two
obeys the spectral addition rule (Eq. \ref{eq:S_xx}), $g^2(\omega)$ and
$N(\omega)$ have the same functional dependence on $\omega$. Hence,
$g^2(\omega)/N(\omega)$ is independent of $\omega$, which means that this motif
can transmit signals at all frequencies with the same fidelity.

For both motifs II and III the mutual information between trajectories does not
depend on the deactivation rate $\mu$ of the read-out component $X$;
$g^2(\omega)$ and $N(\omega)$ depend in the same way on $\mu$. The information
on the input trajectory $s(t)$ is encoded solely in the statistics of the
{\em production} events of X; decays of X occur independently of $S$ and hence
provide no new information about $S$. These observations may suggest that if an
input signal is detected via one of these  motifs, the deactivation rate of $X$
is not important. However, if the information encoded in $X$ needs to be
transmitted to a downstream pathway, then this transmission rate will in general
depend on $\mu$.

Recently, Endres and Wingreen \cite{Endres08} have argued that detection motif
II is superior to motif III in measuring average concentrations. Our analysis
shows that motif II can also more reliably transmit information in time-varying
signals, due to the more accurate transmission of high frequency components of
the input.

{\em Bacterial chemotaxis}---A classical example of a biological system in which
not only the instantaneous value of the input signal is important, but also its
history, is the chemotaxis system of {\em Escherichia coli} \cite{Block83}. The
messenger protein CheY is phosphorylated (${\rm CheY_p}$) by the kinase CheA and
dephosphorylated by the phosphatase CheZ. The kinase activity is rapidly
inhibited by receptor-ligand binding, allowing the system to respond to changes
in ligand concentration on short time scales. Receptor methylation slowly
counteracts the effect of ligand binding on CheA activity, allowing the system
to adapt to changes in ligand concentration on longer time scales. An open
question is how this network processes the ligand signal in the presence of
noise \cite{Andrews06}. Here, we study how reliably the chemotaxis network can
transmit information in time-varying input signals.

\begin{figure}[tb]
\includegraphics{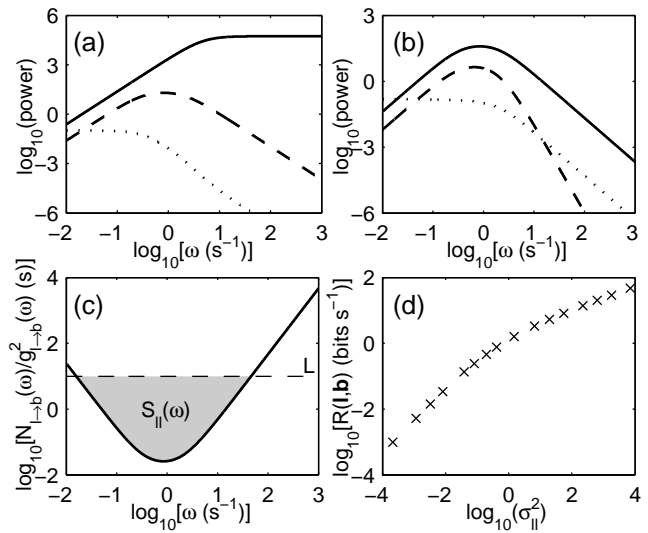}
\caption{Information transmission in the {\em E. coli} chemotaxis network. The
	network gain $g^2(\omega)$ (dashed line), noise $N(\omega)$ (dotted
	line) and $g^2(\omega)/N(\omega)$ (full line) are shown between (a)
	ligand and ${\rm CheY_p}$ concentrations, and (b) ligand and motor bias.
	(c) Water filling approach for the optimal input signal \cite{Fano}. 
	The optimal power spectrum, subject to a total power constraint 
	$\sigma_{ll}^2=\int_{-\infty}^{\infty}S_{ll}(\omega)d\omega$, is given
	by $S_{ll}(\omega)=L-N_{l\to b}(\omega)/g^2_{l\to b}(\omega)$, with $L$
	chosen such that the shaded area matches $\sigma_{ll}^2$. (d)
	$R(\mathbf{l},\mathbf{b})$ evaluated numerically for different
	$\sigma_{ll}^2$ values when the corresponding optimal input power
	spectrum is chosen. The following parameter values were used, estimated
	from \cite{Tu08_2,Emonet08}: $\alpha=2.7$, $\beta=1.3$, 
	$\tau_m=8{\rm s}$, $\langle\eta_m^2\rangle=10^{-4}{\rm	s}^{-1}$,
	$\gamma=8{\rm s}^{-1}$, $\tau_z=0.5{\rm s}$, 
	$\langle\eta_y^2\rangle=0.002{\rm s}^{-1}$, $k=1{\rm s}^{-1}$,
	$\tau_b=0.5{\rm s}$, $\langle\eta_b^2\rangle=0.5{\rm s}^{-1}$.}
\label{fig:chemo}
\label{fig:water_filling}
\end{figure}

Recently, Tu {\em et al.} have shown that a minimal model can accurately
describe the response of the chemotaxis system to a wide range of time-varying
input signals \cite{Tu08_2}. In this model it is assumed that receptor-ligand
binding and the kinase response are much faster than ${\rm CheY_p}$
dephosphorylation and receptor (de)methylation; hence the kinase activity is in
quasi-steady-state. Linearizing around steady-state, we obtain the following
model \cite{Tu08_2}:
\begin{eqnarray}
a(t)&=&\alpha m(t)-\beta l(t)\\ 
\frac{dm}{dt}&=&-\frac{a(t)}{\tau_m}+\eta_m(t) \\
\frac{dy}{dt}&=&\gamma a(t)-\frac{y(t)}{\tau_z}+\eta_y(t).
\end{eqnarray}
Here, $a(t)$ and $m(t)$ are, respectively, the deviations of the fraction of
active kinases and the receptor methylation level from their steady-state
values; $l(t)$ and $y(t)$ are the fractional changes in the ligand and ${\rm
CheY_p}$ concentrations relative to steady-state levels; $\tau_m$ and $\tau_z$
are the time scales for receptor (de)methylation and ${\rm CheY_p}$
dephosphorylation, with $\tau_m>\tau_z$; $\eta_m$ and $\eta_y$ are Gaussian
white-noise sources that are independent of one another, and of the ligand
signal: $\langle \eta (t)\rangle = 0$; $\langle \eta(t)\eta(t^\prime)\rangle =
\langle \eta^2 \rangle \delta(t-t^\prime)$; $\langle\eta_m(t)\eta_y(t')\rangle
=\langle\eta(t)l(t') \rangle=0$. The statistics of the input signal are
described by the power spectrum $S_{ll}(\omega)$.  This system obeys the
spectral addition rule (Eq. \ref{eq:S_xx}), and the power spectrum of $y$ is
given by
\begin{equation}
S_{yy}(\omega) = N_{l\to y}(\omega) + g^2_{l\to y}(\omega) S_{ll}(\omega),
\label{eq:S_yy}
\end{equation}
with $N_{l\to y}(\omega)$ and $g^2_{l\to y}(\omega)$ being intrinsic properties
of the chemotaxis network:
\begin{eqnarray}
g_{l\to y}^2(\omega)&=&\frac{\beta^2\gamma^2\omega^2}{(\omega^2+\tau_z^{-2})
	(\omega^2+\alpha^2/\tau_m^2)}, \label{eq:g_y}\\
N_{l\to y}(\omega)&=&\frac{\alpha^2\gamma^2\langle\eta_m^2\rangle+
	(\alpha^2/\tau_m^2+\omega^2)\langle\eta_y^2\rangle}{
	(\omega^2+\tau_z^{-2})(\omega^2+\alpha^2/\tau_m^2)}. \label{eq:N_y}
\end{eqnarray}
Eq. \ref{eq:g_y} shows that the gain is small at low frequencies, due to
adaptation of the kinase activity via receptor methylation \cite{Tu08_2}
(Fig. \ref{fig:chemo}a). This network is therefore unable to respond to
low-frequency variations in the ligand signal. As noted in
\cite{Tu08_2,Emonet08}, the gain also decreases at high frequencies, due to the
time taken for ${\rm CheY_p}$ dephosphorylation by CheZ. However, we see that
the noise also decreases with increasing frequency. In fact, at high frequencies
the methylation dynamics can be ignored, and the dynamics of ${\rm CheY_p}$ are
approximately those of motif III, discussed above; 
$g^2_{l\to y}(\omega)/N_{l\to y}(\omega)$ increases to a constant value showing
that, in contrast to the conclusions of \cite{Tu08_2}, high-frequency signals
can be reliably encoded in the trajectory $y(t)$.

However, the ultimate response of this system is that of the flagellar motor.
Binding of ${\rm CheY_p}$ to the motor increases the tendency of the motor to
switch to the clockwise state, which causes the bacterium to ``tumble'' and
change direction. Assuming that ${\rm CheY_p}$ binding to the motor is fast,
and that the motor response can be linearized, the clockwise bias of the motor
$b(t)$ is determined by
\begin{equation}
\frac{db}{dt}=k y(t)-\frac{b(t)}{\tau_b} +\eta_b(t),
\end{equation}
where $\tau_b$ is the typical motor switching time and $\eta_b$ represents
Gaussian white noise, uncorrelated from $\eta_m$ and $\eta_y$. Applying the
spectral addition rule, the power spectrum of the motor is $S_{bb}(\omega) =
N_{y\to b}(\omega) + g^2_{y\to b}(\omega)S_{yy}(\omega),$ where $N_{y\to
b}(\omega) = \langle \eta_b^2 \rangle/(\omega^2+\tau_b^{-2})$ is the intrinsic
noise of the motor, and $g^2_{y\to b}(\omega)=k^2/(\omega^2+\tau_b^{-2})$ is the
frequency-dependent gain of the motor. Inserting $S_{yy}(\omega)$ of Eq.
\ref{eq:S_yy} into this expression for $S_{bb}(\omega)$, we see that the total
noise added between the ligand and the motor is given by $N_{l\to b}(\omega)
=N_{y\to b}(\omega)+g^2_{y\to b}(\omega) N_{l\to y}(\omega)$, while the overall
gain of the network is $g^2_{l\to b}(\omega)=g^2_{l\to y}(\omega)g^2_{y\to b}
(\omega)$.

Fig. \ref{fig:chemo}b shows that $g^2_{l\to b}(\omega)$ is large at frequencies
$\tau_m^{-1}\lesssim\omega\lesssim\tau_z^{-1}\sim\tau_b^{-1}$, while $N_{l\to
b}(\omega)$ monotonically decreases with increasing frequency. Importantly, at
high frequencies $\omega\gg\tau_z^{-1},\tau_b^{-1}$, the gain $g^2_{l\to
b}(\omega)\sim\omega^{-4}$ since both $g^2_{l\to y}(\omega)$ and $g^2_{y\to
b}(\omega)$ decrease as $\omega^{-2}$, while $N_{l\to b}(\omega)\sim\omega^{-2}$
since the dominant noise contribution is the intrinsic noise of motor switching
$N_{y\to b}(\omega)$. As a result, $g^2_{l\to b}(\omega)/N_{l\to b}(\omega)$
scales as $\omega^{-2}$ for high frequencies. Hence, while high-frequency
fluctuations in $l(t)$ are reliably encoded in the trajectory $y(t)$, this
information is not propagated to the motor. In essence, the high-frequency
variations of $l(t)$ are filtered by the slow dynamics of ${\rm CheY_p}$
dephosphorylation and motor switching, and are therefore masked by the
inevitable intrinsic noise of motor switching.

The goal of the chemotaxis network is to determine whether the ligand
concentration has increased or decreased. This binary decision has to be made on
the timescale of a motor switching event, which means that the network should
recover at least one bit of information from the input trajectory over this
timescale: $R(\mathbf{l},\mathbf{b})>1{\rm bit}/\tau_b=2{\rm bits\ s}^{-1}$. Our
results allow us to predict the input power spectrum $S_{ll}(\omega)$ that
maximizes $R(\mathbf{l},\mathbf{b})$ for a given power constraint
$\sigma^2_{ll}$ (see Fig. \ref{fig:water_filling}c), which is peaked around
$\omega\approx1{\rm s}^{-1}$. Fig. \ref{fig:water_filling}d shows the
corresponding optimal information rate as a function of $\sigma^2_{ll}$, and
suggests that to achieve $R(\mathbf{l},\mathbf{b})>2{\rm bits\ s}^{-1}$ a
signal variance of at least $\sigma^2_{ll}\approx 2.5$ is required. The
predicted form of the gain-to-noise ratio and the optimal input power spectrum
could be tested by exposing {\em E. coli} cells to oscillating stimuli with
different frequencies, for example in a microfluidic device, and measuring the
(cross) power spectra of the motor bias and the stimulus.

The input signal that a bacterium perceives depends not only on the
spatio-temporal correlations of the ligand concentration in the environment but
also on its swimming behavior, which in turn depends on the input signal itself:
as Fig. \ref{fig:water_filling}b shows, {\em E. coli} is unable to reliably
respond to high- ($\omega\gg\tau_z^{-1},\tau_b^{-1}$) or low-frequency
($\omega\ll\tau_m^{-1}$) stimuli. This means that, in order to find food, {\em
E. coli} should swim neither too slowly nor too fast. Specifically, our
predicted optimal input spectrum suggests that chemotaxis is most efficient when
the spatio-temporal correlations of the ligand and the swimming speed of the
bacterium are matched to give a typical frequency of the ligand signal of about
$\omega \approx 1{\rm s}^{-1}$. Further work is needed to study whether nature
has optimized this feedback between swimming and signaling, and to explore the
naturally occurring chemoattractant distributions that {\em E. coli} would
experience.

We thank Martin Howard and Sorin T\u{a}nase-Nicola for a critical reading of the
manuscript. This work is supported by FOM/NWO.

\end{document}